\documentclass[twocolumn,english]{revtex4}
\usepackage{color}
\usepackage{bm}
\usepackage{amsthm}
\usepackage{amsmath}
\usepackage{esint}
\usepackage{graphicx,epsf}
\usepackage{dcolumn}
\usepackage{amsfonts}
\usepackage{mathrsfs}
\usepackage{subfig}

\makeatletter
\begin{document}

\title{Effects of radial electric field on ion temperature gradient driven mode stability}

\author{Ningfei Chen, Hanyuan Hu, Xiangyu Zhang, Shizhao Wei and Zhiyong Qiu\footnote{E-mail: zqiu@zju.edu.cn}}

\affiliation{Institute for Fusion Theory and Simulation and Department of Physics, Zhejiang University, Hangzhou 310027, China}

\date{\today}
\begin{abstract}
The local  stability of  ion-temperature gradient driven mode (ITG) in the presence of a given radial electric field is investigated
using gyrokinetic theory and ballooning mode formalism with toroidal effect accounted for. It is found that, zero frequency radial electric field induced poloidal rotation can significantly stabilize ITG, while the associated density perturbation has little effect on ITG stability due to the modification of finite-orbit-width effect. However, the parallel mode structure is slightly affected due to the evenly symmetric density modulation of ZFZF.
\end{abstract}
\pacs{}

\maketitle

\section{Introduction}

Drift waves (DWs) turbulence \cite{HortonRMP1999}, driven by free energy associated with plasma pressure gradients, are considered as candidates for inducing anomalous plasma transport and degradation of confinement in magnetically-confined fusion (MCF) devices. Ion-temperature gradient driven mode (ITG) is one of the most intensively studied DWs due to its potential role in causing anomalous ion thermal transport, which is much concerned in future fusion reactors. ITG has two branches, i.e., a slab branch by the coupling of ion parallel compression and diamagnetic drift, and a toroidal branch by the coupling of diamagnetic drift with the unfavored curvature in the weak field side \cite{CZChengPoF1980,LChenPoFB1991}. In-depth understanding of the mechanisms for ITG linear stability, nonlinear evolution and eventual saturation, is needed for quantitative understanding of plasma confinement in future tokamaks. Excitation of zonal flows (ZFs), is considered as an important route for ITG self-regulation, and the regulation is achieved via nonlinear excitation of ZFs by ITG via modulation instability as ITG amplitude exceeds the threshold induced by frequency mismatch, which in turn, scatters ITG into the linearly stable short radial wavelength regime \cite{LChenPoP2000,ZoncaEPL2008}.

ZFs are typically meso-scale radial corrugations with toroidally symmetric ($n=0$), and predominantly poloidally symmetric ($m\simeq0$) scalar potential fluctuation, and consist of zero-frequency ZF (ZFZF) \cite{DiamondPPCF2005} and its finite frequency counter-part, geodesic acoustic mode (GAM) \cite{WinsorPoF1968,ZoncaEPL2008}. Here, $m$/$n$ are the poloidal/toroidal mode numbers of the torus. The nonlinear interaction of ITG with ZFs are observed in experiments \cite{KZhaoPRL2006,GConwayPPCF2008,TLanPoP2008,WLZhongNF2015,MelnikovNF2017}, as well as in large scale simulations \cite{ZLinSCI1998,HahmPPCF2000,LiuFengPoP2010,XLiaoPoP2016} where ITG and the associated transport are suppressed,
and it is also found the threshold on pressure gradient on ITG stability is up-shifted as nonlinear effects are taken into account \cite{DimitsPoP2000}. Furthermore, the radial electric field $E_{r}$ associated with large-scale mean flow, as well as its gradient, is also observed to be related to turbulence suppression and confinement improvement, and possibly the formation of transport barrier,  as well as transition from low- to high-confinement regime.

Several models are proposed to investigate the mechanism of ITG suppression by ZFs, among which are $E\times B$ shear effect and radial envelope modulation. In the $E\times B$ shearing model, a two-point nonlinear theory is proposed to understand flow shear suppression of ITG turbulence in cylindrical and toroidal geometry, and both high/low-frequency component of  the radial electric field is considered \cite{HahmPoP1995,HahmPoP1999}. It is found that   a significant reduction of turbulence activity occurs when the shearing rate $\omega_S$, which is proportional to $\partial(v_\theta/r)/\partial_r$, with $v_\theta=E_r/B$ being the radial electric field induced $E\times B$ drift velocity, exceeds the decorrelation rate of the ambient turbulence. Additionally, compared to the low-frequency $E_r$, the high-frequency component of $E_r$ typically plays less significant role in turbulence suppression due to its oscillating nature. On the other hand, the radial envelope modulation model may be referred to as the scattering process, i.e., the potential well associated with $E_r$ scatters turbulence into linearly stable short radial wavelength domain. The thresholds on  the ITG amplitude for  the ZFZF and/or GAM generation, however, are comparable to each other \cite{ZoncaEPL2008}. Global descriptions are needed in both models, which   have previously been based on the local description \cite{RomanelliPoFB1993}. More specifically, the modulational instability describing nonlinear interaction between ITG and ZFZF requires  the ITG linear dispersion relation with finite-$\theta_k$ modification to the local one. Here, $\theta_k\equiv k_r/(n\partial_r q)$ with $k_r$ being the radial envelope wavenumber, and  $q$ being the safety factor. Investigation of the  interaction between turbulence and radial electric field in terms of  the parallel mode structure is not found in  the  literature  to date. Typically, this issue is treated in  the gyrokinetic framework with toroidal effects neglected \cite{ArtunPoFB1992,GaoPoP2004}.

In this work, a local model is proposed to figure out the local properties of turbulence suppression by given $E_r$ and study the ``linear'' stability of ITG in the presence of $E_r$. Here, ``local" means the ITG eigenmode equation
 is solved along the magnetic field lines, with toroidal effects and parallel compression properly accounted for, while physics associated with radial envelope is neglected systematically. Technically,  this is achieved by deriving an ITG governing eigenmode equation in the existence of the radial electric field induced density modulation as well as poloidal rotation, which is then solved   in ballooning space  for the ITG local dispersion relation and mode structure \cite{ConnorPRSL1979,Taylor1977SC}.    For simplicity of discussion and  because of  the temporal scale separation, ZFZF-type $E_r$  on ITG local stability is investigated as an example  \cite{LChenPoP2000}. Analogous to the ZFZF, finite-frequency GAM effects on ITG stability may be treated similarly if time scale separation between ITG and GAM is satisfied, which is typically the case as we discuss in the final section.   Finally, our model may also shed light on  turbulence suppression by  the mean  flow, whose mechanism is not yet fully understood.

The rest of the paper   is organized as follows. In section \ref{sec:2}, the ITG eigenmode equation in the  presence of  a given radial electric field   is derived using gyrokinetic theory and ballooning mode representation. In section \ref{sec:3}, the ITG stability   is investigated assuming a radial electric field with zero frequency, i.e., that of ZFZF, in both  the short- and long-wavelength limit. Summary and discussions are given in Section \ref{sec:4}.

\section{General formalism\label{sec:2}}

For simplicity of discussion while focusing on the main scope of the
present paper, we consider a tokamak with axisymmetric concentric
circular magnetic surface and straight field line, and a left-handed
coordinate $\left(r,\theta,\phi\right)$ is adopted, with $r$, $\theta$
and $\phi$ being the minor radius, poloidal and toroidal angles of
the torus, respectively. The equilibrium magnetic field is given as
$\mathbf{B}=B_{0}[(1-\epsilon\cos\theta)\mathbf{e_{\phi}}+\epsilon\mathbf{e_{\theta}}/q]$,
where $\epsilon\equiv r/R$ is the inverse aspect ratio, $R$ is the on-axis
major radius and $q\equiv rB_{\phi}/(RB_{\theta})$.
ITG  modes generally have ballooning structure with high
mode numbers, and the characteristic scale of equilibrium profile is generally much larger
than  the distance between neighbouring mode rational surfaces. Consequently,
the perturbed quantity can be expressed as
\begin{align}
\delta\phi & =e^{in\phi-i\omega t}\underset{j}{\sum}\hat{\Phi}\left(s-j\right)e^{-i\left(m_{0}+j\right)\theta}.\label{eq:1}
\end{align}
Here, $s\equiv(r-r_{0})/\Delta r=nq-m_{0}$, $r_{0}$ denotes the reference rational surface with $nq(r_0)-m_0=0$, $\Delta r=1/(n\partial q/\partial r)$
is the distance between neighboring mode rational surfaces, and
$\left|j\right|\ll m_{0}$ is an integer.

The   gyrokinetic equation \cite{TaylorPP1968} is used
to investigate the stability of ITG turbulence   in the presence of a given radial electric
field. Following Ref. \cite{LChenPoFB1991}, we take the flat density
gradient limit to focus on effects of ion temperature gradient, i.e., assuming
$\eta_{i}=L_{ni}/L_{Ti}\rightarrow\infty$, with $L_{ni}=-n_{i}/(\partial n_{i}/\partial r)$
and $L_{Ti}=-T_{i}/(\partial T_{i}/\partial r)$ being the characteristic scale
length of ion density and temperature nonuniformity, respectively. The gyrokinetic equation
for ion response to ITG can be written as
\begin{align}
\left(\omega-k_{\parallel}v_{\parallel}+\omega_{D}+  k_{\theta} v_{\theta} \right)\delta H_{I}= & \dfrac{e}{T_{i}}J_{0}\left(\omega+\omega_{*i}^{T}\right)F_{0i}\delta\phi_{I}\nonumber \\
 -i\frac{c}{B_0}\mathbf{b}\times\nabla\delta\phi_I\cdot\nabla\delta H_E.\label{eq:2}
\end{align}
Here, $k_{\parallel}\equiv\left(nq-m\right)/\left(qR\right)$ is the
parallel wavenumber, $\omega_{D}\equiv2\omega_{d}C\left(x_{\perp}^{2}/2+x_{\parallel}^{2}\right)$
is the magnetic drift frequency, with $\omega_{d}=k_{\theta}cT_{i}/(eBR)$,
$x_{\perp}=v_{\perp}/v_{ti}$ and $x_{\parallel}=v_{\parallel}/v_{ti}$
being the ion perpendicular/parallel velocities normalized by thermal
velocity $v_{ti}=\sqrt{2T_{i}/m_{i}}$, respectively.  $C=\cos\theta-k_{r}\sin\theta/k_{\theta}$ is related to the curvature
with $k_{r}$ and $k_{\theta}=m_{0}/r_{0}$ being the
radial/poloidal mode numbers. $\delta H_{I}$ is the nonadiabatic ion
response to ITG, $J_{0}\left(k_{\perp}\rho_{i}\right)$ is the Bessel function
of zero-index accounting for Finite Larmor radius (FLR) effects,
$\rho_{s}=mv_{\perp,s}c/(eB)$ is the Larmor radius of species $s$, $F_{0i}$ is the equilibrium ion distribution function, and $\omega_{*i}^{T}=\omega_{*Ti}\left(x_{\perp}^{2}+x_{\parallel}^{2}-3/2\right)$
is the ion diamagnetic frequency in the flat density limit, with $\omega_{* Ti}=k_{\theta}cT_{i}/(eBL_{Ti})$.
Furthermore, the last term on the left hand side accounts for the Doppler shift   from the radial electric field induced poloidal rotation, with $v_{\theta}\equiv -c\mathbf{b}\times\partial_r\delta\phi_E/B_0$, while the last term on the right hand side represents the  perturbed diamagnetic term associated with the  density perturbation induced by the  radial electric field \cite{LChenEPL2014}.  The last term, in fact,  can be combined with the term proportional to $\omega^T_{*i}$ , considering $F_0 + \delta H_E$   to be the renormalized equilibrium in the existence of the radial electric field.
It is   worth noting that the two additional terms, i.e., radial electric field induced poloidal rotation of ITG $k_{\theta}v_\theta\delta H_{I}$, and  variation along the magnetic field line induced by  the density perturbation associated with the radial electric field $\propto \delta\phi_{I}\delta H_{E}$, can also be obtained from the perpendicular nonlinear term in nonlinear gyrokinetic equation \cite{FriemanPoF1982}, and thus, will be called ``nonlinear terms" in the following discussion for convenience, though the radial electric field can also originate from linear effects, such as large scale mean flow.  Here, subscripts ``E" and ``I" represent quantities associated with radial electric field and ITG, respectively. The dispersion relation can be derived from charge quasi-neutrality condition
\begin{equation}
\dfrac{eN_{0}\delta\phi}{T_{e}}+\left\langle \delta H_{e}J_{0}\right\rangle =-\dfrac{eN_{0}\delta\phi}{T_{i}}+\left\langle \delta H_{i}J_{0}\right\rangle ,\label{eq:3}
\end{equation}
with $eN_{0}\delta\phi/T_{e}$ and $-eN_{0}\delta\phi/T_{i}$ being
adiabatic responses of electron and ion, respectively, and $\left\langle \cdots\right\rangle$
representing velocity space integration. The derivation follows closely
the procedure of Ref. \cite{GuzdarPoF1983}. For typical ITG fluctuation with
$k_{\parallel}v_{\parallel,e}\gg\omega\sim\omega_{*i}\gg\omega_{d}$,
$k_{\parallel}v_{\parallel,i}$, electrons responde adiabatically, i.e., $\delta H_{I,e}=0$. The nonadiabatic ion response
can be derived as
\begin{eqnarray}
\delta H_{I,i}\approx\dfrac{\Lambda}{\omega}\left[-\left(\dfrac{e}{T_{e}}+\dfrac{e}{T_{i}}\right)F_{0}\delta\phi_{E}+\delta H_{E}\right]\delta\phi_{I}\nonumber \\
+\dfrac{e}{T_{i}}J_{0}F_{0}\left(1+\dfrac{\omega_{*i}^{T}}{\omega}\right)\left(1+\dfrac{k_{\parallel}v_{\parallel}}{\omega}+ \dfrac{k_{\parallel}^{2}v_{\parallel}^{2}}{\omega^{2}}-\dfrac{\omega_{D}}{\omega}\right)\delta\phi_{I}.\label{eq:4}
\end{eqnarray}
Here, $\Lambda\equiv ick_{r}k_{\theta}/B_{0}$, and will be used in the rest of the paper. The two terms in first bracket of equation (\ref{eq:4}) are the formal nonlinear terms, and represent the effects associated, respectively,
with  the potential and density fluctuation of $E_{r}$.   The quasi-neutrality condition of ITG is applied to simplify the first term. Substituting the ion and electron response into quasi-neutrality condition (\ref{eq:3}), one then   has the ITG WKB dispersion relation
\begin{eqnarray}
 & \left\{ \dfrac{1}{\tau\left(1+\omega_{*Ti}/\omega\right)} +b_{\perp}-\dfrac{k_{\parallel}^{2}v_{ti}^{2}}{2\omega^{2}}+\dfrac{2\omega_{d}C}{\omega}+\dfrac{\Lambda}{\omega+\omega_{*Ti}}\right.\nonumber \\
 & \left.\times\left[\left(1+\dfrac{1}{\tau}\right)\delta\phi_{E}-\dfrac{T_{i}}{e}\left\langle \dfrac{\delta H_{E,i}}{N_{0}}\right\rangle \right]\right\} \delta\phi=0,\label{eq:5}
\end{eqnarray}
with $b_{\perp}\equiv k_{\perp}^{2}\rho_{i}^{2}/2$, and $k_{\perp}$
being the perpendicular wavenumber. The first four terms of equation (\ref{eq:5})
constitute the linear ITG dispersion relation, with the first three terms being respectively, adiabatic electron response, the FLR effect (polarization) and parallel compressibility, while the forth term  related to magnetic drift peculiar in toroidal configuration, resulting in coupling of neighbouring poloidal harmonics.
The last two terms are nonlinear modifications
due to poloidal rotation and density modulation associated with the radial
electric field, respectively. Noting $k_{\perp}^{2}=k_{\theta}^{2}-\partial^{2}/\partial r^{2}$, the eigenmode equation in real space for $j$-th poloidal harmonics can be derived as
\begin{eqnarray}
\left(b_{\theta}\hat{s}^{2}\dfrac{d^{2}}{dz^{2}} -\dfrac{1}{\tau\left(1+\omega_{*Ti}/\omega\right)}-b_{\theta}+\dfrac{k_{\parallel}^{2}v_{ti}^{2}}{2\omega^{2}}\right)\hat{\Phi}_{z}\nonumber \\
=\dfrac{\omega_{d}}{\omega}\left[\hat{\Phi}_{z+1}+\hat{\Phi}_{z-1}+\hat{s}\dfrac{d}{dz}\left(\hat{\Phi}_{z+1}-\hat{\Phi}_{z-1}\right)\right]\nonumber \\
+\dfrac{\Lambda}{\left(\omega+\omega_{*Ti}\right)}\left[\left(1+\dfrac{1}{\tau}\right)\delta\phi_{E}-\left\langle \dfrac{T_{i}\delta H_{E,i}}{eN_{0}}\right\rangle \right]\hat{\Phi}_{z}.\label{eq:6}
\end{eqnarray}
Here, $\hat{s}\equiv r\left(\partial q/\partial r\right)/q$ is the magnetic shear, $\tau\equiv T_{e}/T_{i}$,
$b_{\theta}\equiv k_{\theta}^{2}\rho_{ti}^{2}/2$, $z\equiv s-j=nq-m$
is the normalized distance to the mode rational surface. The first term on the right-hand side of equation (\ref{eq:6})
comes from the curvature drift induced coupling between neighbouring poloidal harmonics. Moreover,
the term proportional to $\left\langle\delta H_{E,i}\right\rangle$ may also have poloidal-dependence,
and causes additional toroidal coupling. For instance, GAM with $\omega\gg \omega_{tr,i}$,
is characterised by up-down anti-symmetric ($\propto\sin\theta$) density fluctuation \cite{ZoncaEPL2008}, while
ZFZF with $\omega\ll \omega_{tr,i}$, has $\cos\theta$-type density
fluctuation \cite{LChenEPL2014}. Here $\omega_{tr,i}\equiv v_{\parallel,i}/(qR)$ is the circulating ion transit frequency. Equation (\ref{eq:6}) can be analyzed using the ballooning
mode formalism framework \cite{ConnorPRSL1979}, which is accomplished
by taking $\Phi\left(\eta\right)=\int\hat{\Phi}\left(z\right)\exp\left(-i\eta z\right)dz$,
with $\eta$ being the extended poloidal angle along the magnetic
field lines. The ITG eigenmode equation in ballooning space reads
\begin{eqnarray}
 & \dfrac{d^{2}\Phi\left(\eta\right)}{d\eta^{2}}+q^{2}\Omega^{2}b\left(\dfrac{\tau\Omega}{1+\tau\Omega\epsilon_{Ti}^{1/2}}+b\left(1+\hat{s}^{2}\eta^{2}\right)\right.\nonumber \\
 & +\dfrac{2}{\Omega}\left(\cos\eta+\hat{s}\eta\sin\eta\right)+\left(1+\tau\right)\Delta_{E}\delta\phi_{E}\nonumber \\
 & \left.-\tau\Delta_{E}\left\langle \dfrac{T_{i}\delta H_{E,i}}{eN_{0}}\right\rangle \right)\Phi\left(\eta\right)=0,\label{eq:7}
\end{eqnarray}
where $\Omega\equiv\omega/(\tau\sqrt{\omega_{*Ti}\omega_{d}})$,
$b\equiv\tau b_{\theta}/\sqrt{\epsilon_{Ti}}$, $\epsilon_{Ti}\equiv L_{Ti}/R$
and $\Delta_{E}\equiv \Lambda/[\omega_{*Ti}\sqrt{\epsilon_{Ti}}\left(1+\tau\Omega\sqrt{\epsilon_{Ti}}\right)]$. Equation (\ref{eq:7}) is general and can be applied to study the nonlinear modification of any given radial electric field to ITG stability, with the nonlinear modifications by the radial electric field accounted for by the last two terms.  In this work, as a proof of principle demonstration, we will consider ZFZF-type stationary radial electric field, while the effects of energetic particle induced GAM (EGAM)/GAM can be investigated straightforwardly following the same approach if the GAM/EGAM frequency is smaller than ITG growth rate. It is natural to take the dominant $m=0,1$ components of nonadiabatic ion response \cite{ZYQiuPPCF2008},
\begin{eqnarray}
 & \delta H_{E,i}=\dfrac{eF_{0}\overline{\delta\phi_{Z}}}{T_{i}}\left(1+i\lambda\cos\theta\right),\label{eq:8}
\end{eqnarray}
and $m=0$ component of $\delta\phi_{Z}$, i.e., $\overline{\delta\phi_{Z}}$,
where $\left(\overline{\cdots}\right)\equiv\int_{0}^{2\pi}(\cdots) d\theta/2\pi$ represents surface averaged quantity.  The higher order $m=1$ density perturbation of ZFZF is included, to account for its unique role in inducing periodic modification to the ITG eigenmode potential well along the magnetic field line, that determines the condition for ITG stability.  Here, $\lambda=\omega_{Dr}/\omega_{tr,i}$ represents the finite drift orbit width effect, with $\omega_{Dr}=2\omega_{dr}(x_\perp^2/2+x^2_\parallel)$ being the magnetic
drift frequency associated with normal curvature and $\omega_{dr}\equiv k_{r}cT_{i}/(eBR)$.

\section{Effects of zero-frequency $E_{r}$ on ITG linear stability\label{sec:3}}

With  the specified expression of $\delta H_{E,i}$ presented in equation (\ref{eq:8}),
equation  (\ref{eq:7}) can be written as
\begin{eqnarray}
 &  & \dfrac{d^{2}\Phi}{d\eta^{2}}+q^{2}\Omega^{2}b\left(\dfrac{\tau\Omega}{1+\tau\Omega\epsilon_{Ti}^{1/2}}+b\left(1+\hat{s}^{2}\eta^{2}\right)+\Delta_{E}\overline{\delta\phi_{Z}}\right.\nonumber \\
 &  & \left.+\dfrac{2}{\Omega}\left(\cos\eta+\hat{s}\eta\sin\eta\right)+\Delta_{E}'\overline{\delta\phi_{Z}}\cos\eta\right)\Phi=0,\label{eq:9}
\end{eqnarray}
where $\Delta_{E}\overline{\delta\phi_{Z}}$ represents modification
due to the electrostatic potential, and $\Delta_{E}'\overline{\delta\phi_{Z}}\cos\eta$
  results from the $m=1$ density perturbation of ZFZF, with $\Delta_{E}'\equiv -i\tau A\left(2\omega_{dr}/\omega_{t,i}\right)\Delta_{E}$ and $A\equiv\left\langle F_0(x_\perp^2/2+x_\parallel^2)/x_\parallel\right\rangle$. Note that, the velocity space integral $A$ vanishes for typical Maxwellian distribution, but finite value for non-even symmetric distribution, e.g., shifted-Maxwellian distribution with non-zero average parallel velocity due to auxiliary current drive.  Equation (\ref{eq:9}) will be investigated in both  the short- and long- wavelength limits, corresponding to strong and moderate ballooning cases, respectively,  as investigated Refs.
\cite{LChenPoFB1991,GuzdarPoF1983}. The two limiting parameter regimes, can be studied by taking $b\sim O(1)$ and $b\ll1$, respectively, as we shown after equation (\ref{eq:10}) that, the mode width in ballooning space, is proportional to $1/\sqrt{b}$ (and thus, $\sim O(\sqrt{b})$ radially).

\subsection{Short-wavelength limit}

\begin{figure}
\subfloat{\includegraphics[scale=0.3]{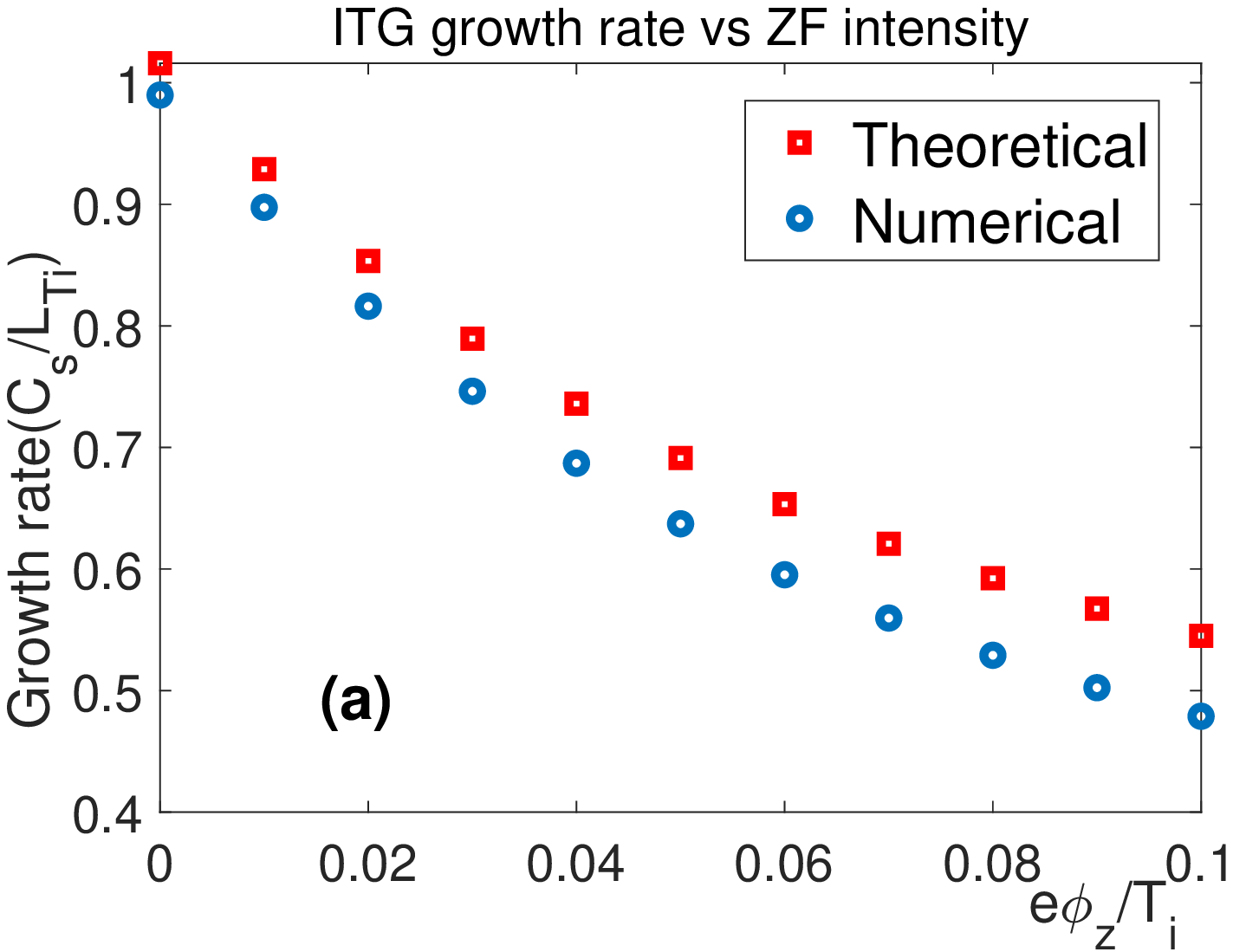}\label{fig:1a}}
\subfloat{\includegraphics[scale=0.3]{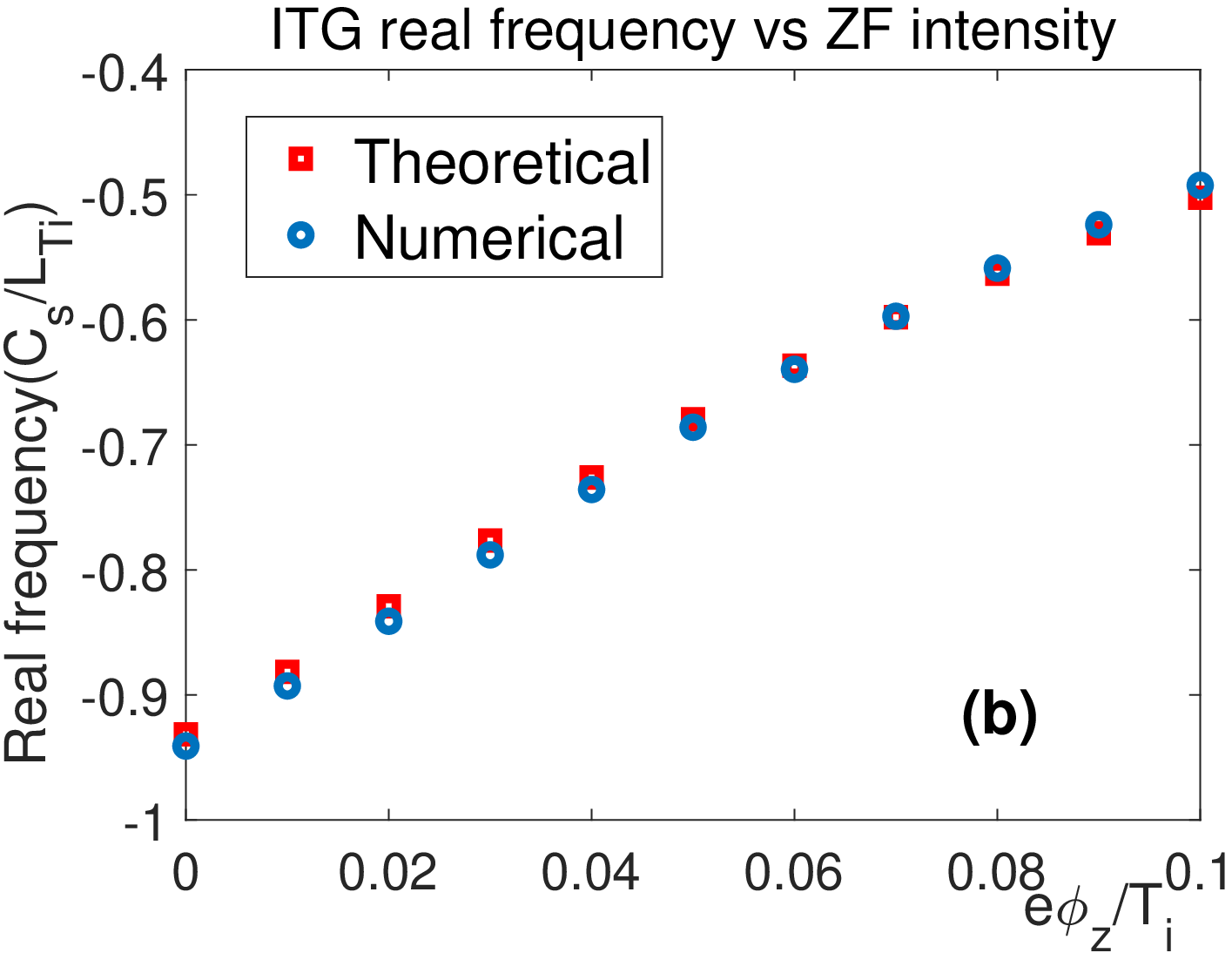}\label{fig:1b}}
\caption{The dependence of normalized growth rate (a) and real frequency (b) of ITG,
which are normalized by $C_{s}/L_{Ti}$, on the normalized ZFZF intensity $e\delta\phi_{Z}/T_{i}$. The squares represent the theoretical
result given by Eq. (\ref{eq:11}) while circles are numerical result
of Eq. (\ref{eq:9}). Here, $C_{s}^{2}\equiv2T_{e}/m_{i}$ is the sound speed, $\epsilon_{Ti}=0.2$, $b=1$ and $A=1$.}
\end{figure}

\begin{figure}
\includegraphics[scale=0.45]{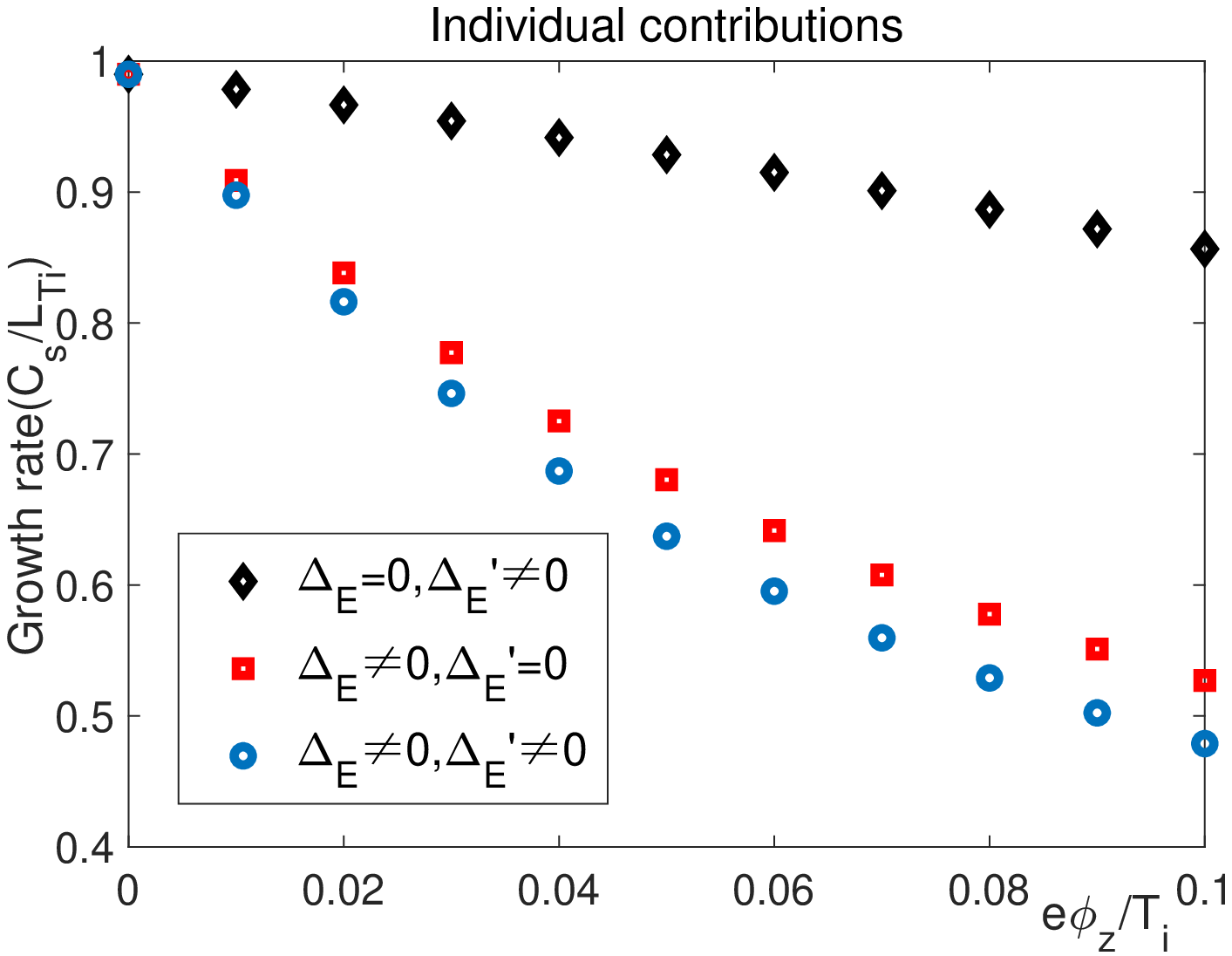}
\caption{\label{fig:2}Numerical result of integrated and separated effects of potential and density fluctuation of ZFZF.}
\end{figure}

\begin{figure}
\includegraphics[scale=0.45]{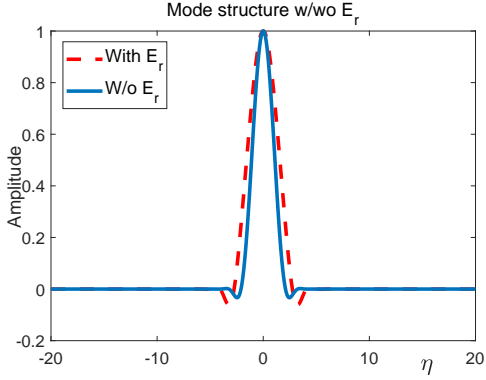}
\caption{\label{fig:3}The mode structure of the lowest eigenmode
when $e\delta\phi_{Z}/T_{i}=0.1$. Here, $\epsilon_{Ti}=0.2$, $b=1$ and $A=1$. }
\end{figure}

In the short-wavelength limit, i.e., $b\sim O\left(1\right)$,
the eigenfunction is strongly localized in ballooning space \cite{LChenPoFB1991}.
Thus, strong coupling approximation can be adopted by taking $\cos\eta\approx1-\eta^{2}/2$ and $\sin\eta\approx\eta$ \cite{Taylor1977SC}. Note that, the  assumption underlying the above strong coupling approximation is that the mode is localized around $\eta=0$, and the introduction of the $\cos\eta$-type periodic modulation does not affect the validity of the assumption due to the even-symmetric of it. The eigenmode equation then becomes
\begin{eqnarray}
 & \dfrac{d^{2}\Phi}{d\eta^{2}}+q^{2}\Omega^{2}b\left(\dfrac{\tau\Omega}{1+\tau\Omega\epsilon_{Ti}^{1/2}}+b+\dfrac{2}{\Omega}+\Delta_{E}\overline{\delta\phi_{Z}}+\Delta_{E}'\overline{\delta\phi_{Z}}\right)\nonumber\\
 & \left.+\left(b\hat{s}^{2}+\dfrac{2\hat{s}-1}{\Omega}-\dfrac{\Delta_{E}'\overline{\delta\phi_{Z}}}{2}\right)\eta^{2}\right)\Phi=0,\label{eq:10}
\end{eqnarray}
which can be rewritten as a standard Weber equation with the most unstable ground eigenmode being given by $\delta\phi=\exp(-\sigma\eta^{2})$ with
\begin{eqnarray*}
\sigma & = & \dfrac{q^2\Omega^2b}{2}\left(\dfrac{\tau \Omega}{1+\tau\Omega\epsilon_{Ti}^{1/2}}+\dfrac{2}{\Omega}+b+\Delta_{E}\overline{\delta\phi_{Z}}+\Delta_{E}'\overline{\delta\phi_{Z}}\right).
\end{eqnarray*}
The half width of the lowest eigenmode in $\eta$ space is proportional to $1/\sqrt{b}$. The corresponding dispersion relation is
\begin{eqnarray}
 & q^{2}\Omega^{2}b\left[\dfrac{\tau\Omega}{1+\tau\Omega\epsilon_{Ti}^{1/2}}+\dfrac{2}{\Omega}+b+\Delta_{E}\overline{\delta\phi_{Z}}+\Delta_E'\overline{\delta\phi_{Z}}\right]^{2}\nonumber \\
 & +\left(b\hat{s}^{2}+\dfrac{2\hat{s}-1}{\Omega}-\dfrac{\Delta_E'\overline{\delta\phi_{Z}}}{2}\right)=0.\label{eq:11}
\end{eqnarray}
The dispersion relation is similar to corresponding linear result \cite{LChenPoFB1991}, except terms proportional to $\Delta_E$ and $\Delta_E'$ originate from the contribution of radial electric field induced poloidal rotation and density
fluctuation, respectively. The dependence of ITG growth rate and real frequency on the radial electric field are solved from the theoretical dispersion relation equation  (\ref{eq:11}), which are then compared with the numerical solution of equation  (\ref{eq:9}), and good agreement between analytical and numerical results are obtained, as shown in Fig. \ref{fig:1a} and \ref{fig:1b}, respectively. It is found in equation (\ref{eq:11}) that the $m=1$ density perturbation is of order $k_{r}\rho_{i}q$ compared to $m=0$ potential fluctuation. Thus, it is obvious that $m=1$ density fluctuation affects the dispersion relation slightly. The ITG growth rate decreases significantly with increasing $\delta\phi_{Z}$. We then analyze the contribution of the radial electric field induced poloidal rotation and density modulation on ITG stability, by turning off the corresponding terms in equation (\ref{eq:9}). It is shown in Fig. \ref{fig:2} that, when the $E_r$ induced poloidal rotation is kept while the density perturbation is turned off, the ITG growth rate is almost the same as that with both effects properly accounted for; while as only the $E_r$ induced density perturbation is kept, the ITG growth rate is slightly affected by the scalar potential. We thus conclude that the reduction of the growth rate is mainly due to the potential fluctuation (poloidal rotation). Besides, it is found that the ITG growth rate is of order $C_{s}/L_{T}$, which is much larger than characteristic GAM/EGAM frequency $\sim C_{s}/R$, hence our analysis can also be applied to modulation of ITG by EGAM/GAM. The mode structure is also shown in Fig. \ref{fig:3}, and it is clearly seen that the mode structure peaks at $\eta=0$ and the even symmetry is  not  broken, resulting from the   even-symmetric period modulation introduced by the  density fluctuation of  ZFZF  ($\propto  \cos\theta$) as denoted by $\Delta'_E$. We note that, equation (\ref{eq:9}) can be further simplified, by substituting the quasi-neutrality condition of ZFZF into equation (\ref{eq:7}) to replace the last term proportional to $\langle \delta H_{E,i}\rangle$. This process will introduce $O(k^2_r\rho^2_i)$ uncertainty since  it is $\langle J_0\delta H_{E,i}\rangle$ in the quasi-neutrality condition instead of the $\langle \delta H_{E,i}\rangle$ in equation (\ref{eq:7}). In this case, effects induced by the $m=1$ density modulation of ZFZF cannot be larger than $O(k^2_r\rho^2_i)$, and is thus weak, as shown by our numerical results.

\subsection{Long-wavelength limit}

\begin{figure}
\subfloat{\includegraphics[scale=0.3]{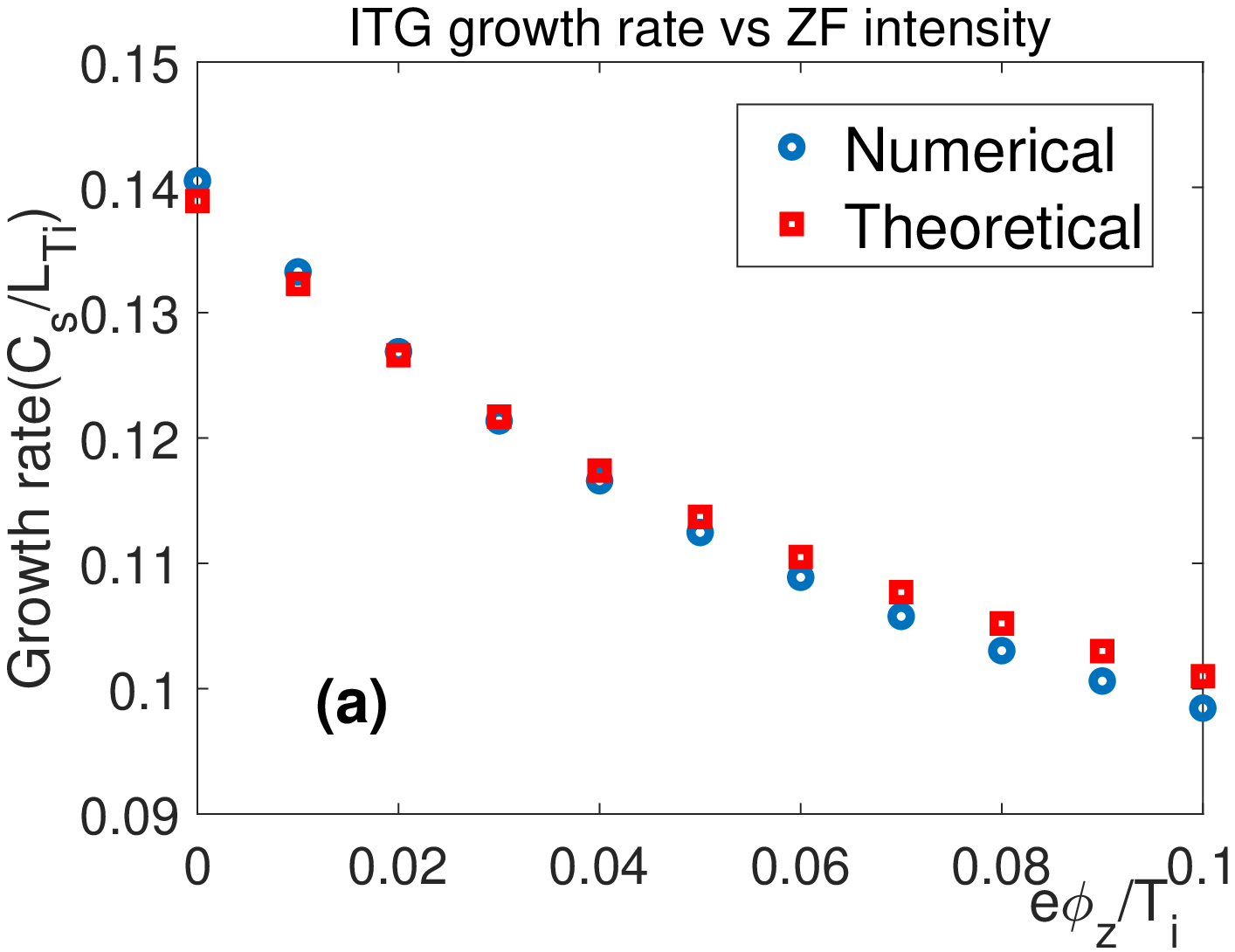}}
\subfloat{\includegraphics[scale=0.3]{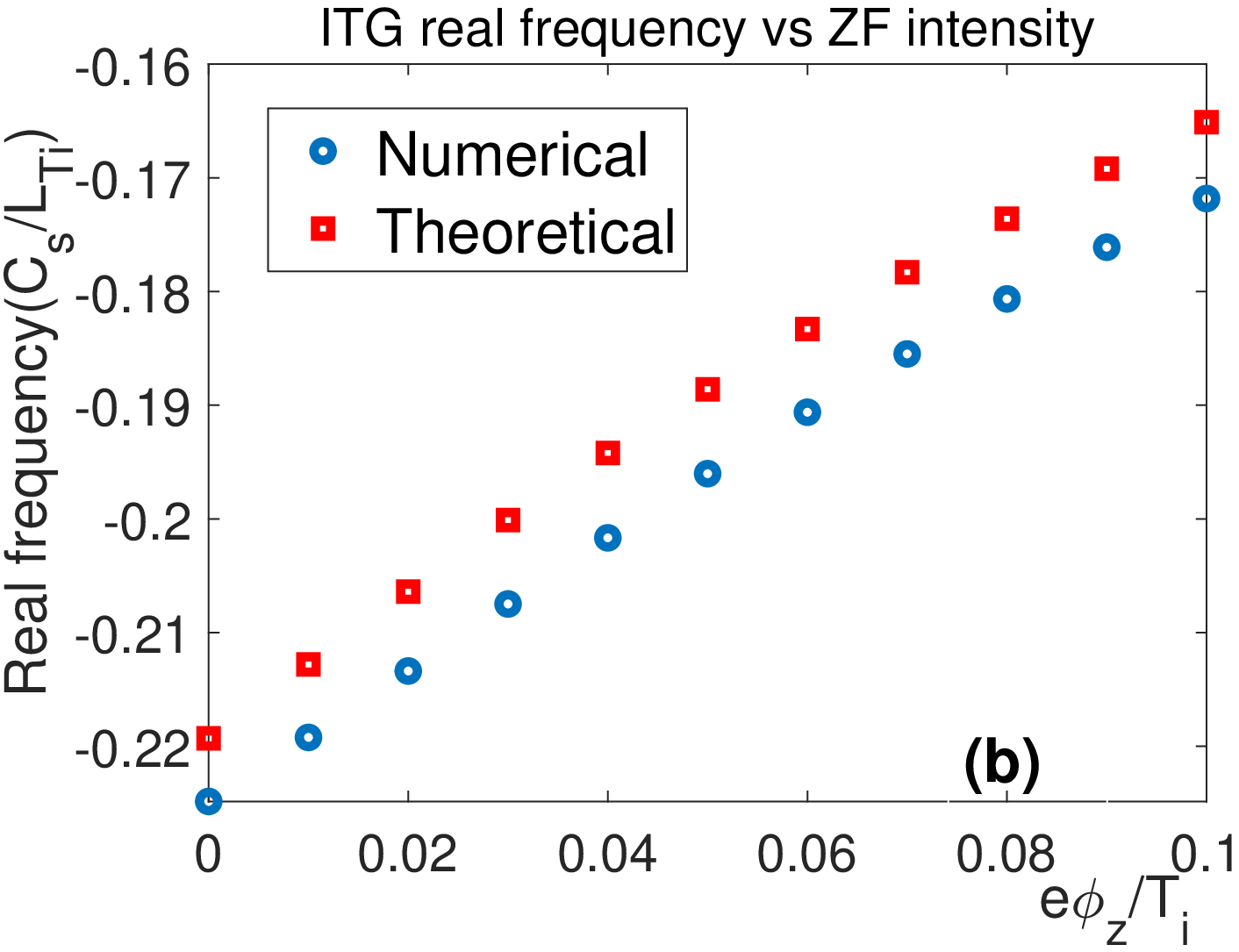}}
\caption{\label{fig:4}The growth rate (a) and real frequency (b) of ITG versus the normalized ZFZF intensity $e\delta\phi_{Z}/T_{i}$. The squares represent the analytical result given by Eq. (\ref{eq:14}) while circles represent numerical result of Eq. (\ref{eq:9}). Here, $\epsilon_{Ti}=0.2$, $b=0.01$ and $A=1$.}
\end{figure}

\begin{figure}
\includegraphics[scale=0.45]{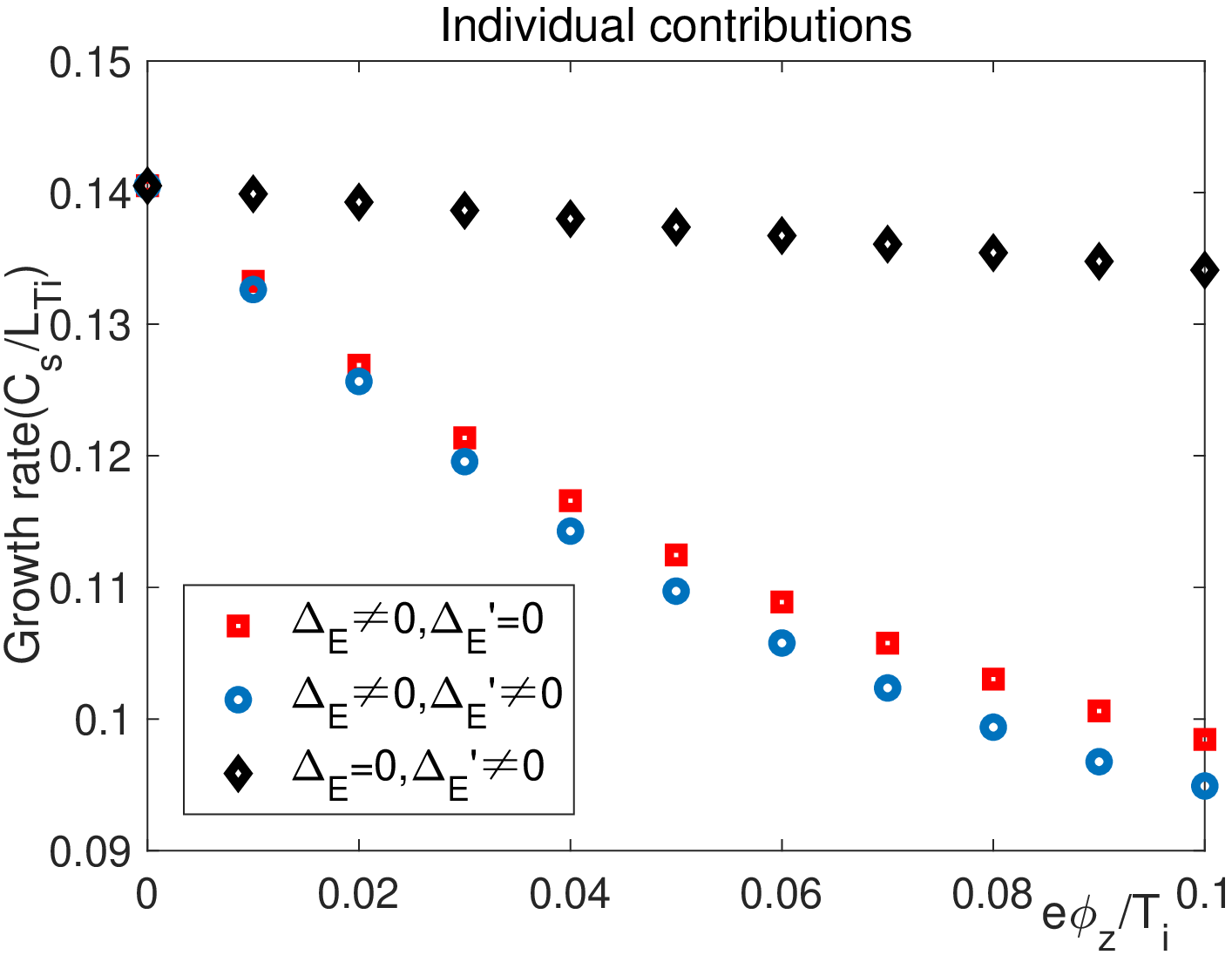}
\caption{\label{fig:5}Numerical result of integrated and separated effects of ZFZF potential and density fluctuation.}
\end{figure}

\begin{figure}
\includegraphics[scale=0.45]{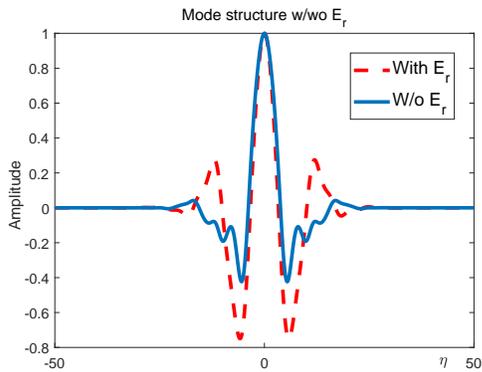}
\caption{\label{fig:6}The mode structure of the most unstable
mode when $e\delta\phi_{Z}/T_{i}=0.1$. Here, $\epsilon_{Ti}=0.2$, $b=0.01$ and $A=1$. }
\end{figure}

For typical tokamak plasmas,  strong coupling approximation is usually a crude constraint. In
more general cases, $b\ll1$ (long-wavelength limit) is satisfied, and strong coupling approximation
no longer holds. In the long-wavelength limit, there are two branches,
i.e., toroidal branch and slab branch. We are more concerned about
the toroidal branch \cite{LChenPoFB1991},
which is characterized by fast variation over connection length scale
($\eta\sim O\left(1\right)$) and a superimposed slowly varying envelope
over secular scale. The self-consistent ordering is given by balancing
parallel compressibility and adiabatic electron response, which results
in $\Omega=O\left(b^{-1/3}\right)$. Taking $\Phi\left(\eta\right)=C_{0}\left(\eta_1\right)\cos\eta/2+S_{0}\left(\eta_1\right)\sin\eta/2$ with $\eta_{1}\equiv\hat{\epsilon}\eta$, and
$\hat{\epsilon}=b^{1/3}$ denoting slow variation in $\eta$, the eigenmode equations can be derived from vanishing
coefficients of $\sin\eta/2$ and $\cos\eta/2$:
\begin{eqnarray}
 & \dfrac{dS_{0}}{d\eta_{1}}+\left[\dfrac{b^{2/3}q^{2}\Omega^{3}\tau}{1+\tau\Omega\epsilon_{Ti}^{1/2}}-\dfrac{1}{4b^{1/3}}+q^{2}\Omega^{2}b^{2/3}\Delta_{E}\overline{\delta\phi_{Z}}\right.\nonumber\\
 & \left.+\dfrac{1}{2}q^{2}\Omega^{2}b^{2/3}\Delta_{E}'\overline{\delta\phi_{Z}}\right]C_{0}+q^{2}\Omega b^{1/3}\hat{s}\eta_{1}S_{0}=0, \label{eq:12}\\
 & \dfrac{dC_{0}}{d\eta_{1}}-\left[\dfrac{b^{2/3}q^{2}\Omega^{3}\tau}{1+\tau\Omega\epsilon_{Ti}^{1/2}}-\dfrac{1}{4b^{1/3}}+q^{2}\Omega^{2}b^{2/3}\Delta_{E}\overline{\delta\phi_{Z}}\right.\nonumber\\
 & \left.+\dfrac{1}{2}q^{2}\Omega^{2}b^{2/3}\Delta_{E}'\overline{\delta\phi_{Z}}\right]S_{0}-q^{2}\Omega b^{1/3}\hat{s}\eta_{1}C_{0}=0. \label{eq:13}
\end{eqnarray}
Equations (\ref{eq:12}) and (\ref{eq:13}) can be cast into a Weber equation for $C_{0}$ and $S_{0}$.
The dispersion relation for the most unstable ground eigenstate is
\begin{eqnarray}
\Omega^{3}+\dfrac{\Lambda\overline{\delta\phi_{Z}}}{\tau\omega_{*T}\epsilon_{Ti}^{1/2}}\left(1-\dfrac{i\tau Ak_{r}\rho_{i}q}{2}\right)\Omega^{2}-\dfrac{\epsilon_{Ti}^{1/2}}{4bq^{2}}\Omega&=&\dfrac{1}{4bq^{2}\tau}.\nonumber\\
\label{eq:14}
\end{eqnarray}
Here, terms in the bracket of Eq. (\ref{eq:14}) comes from the $m=0$ component of radial electric field induced poloidal rotation and $m=1$ density modulation, while other terms originate from linear dispersion relation \cite{LChenPoFB1991}. It is noteworthy that the $m=1$ component of density perturbation enters and affects the ITG dispersion relation, while the $\sin\theta$-type density modulation of EGAM/GAM has no influence on the dispersion relation, possibly due to the odd symmetry of the density modulation of GAM/EGAM with $\omega_G\gtrsim \omega_{tr,i}$. The dependence of ITG growth rate and real frequency on scalar potential of the radial electric field are solved from the theoretical dispersion relation, which are then compared with the numerical solution of Eq. (\ref{eq:9}), and good agreement are obtained, as shown in Fig. \ref{fig:4}. An artificially small $b=0.01$ is adopted to separate different scales,   although this is not the most relevant parameter regime for ITG stability. As shown in Fig. \ref{fig:5}, the $E_r$ induced poloidal rotation is the main reason for the reduction of the ITG growth rate, as clarified by our theoretical analysis; while its density perturbation has weak effect on ITG stability. The mode structure is shown in Fig. \ref{fig:6}, which is wider than that in the short-wavelength limit, as demonstrated by our analysis.

\section{Conclusion and Discussion\label{sec:4}}

In this paper, a governing equation is formulated to investigate the  ITG ``linear" stability in the presence of a given radial electric field, using gyrokinetic equation and ballooning mode representation. The effects of the radial electric field on ITG linear stability consist of $E_{r}$-induced poloidal rotation and density fluctuation, and their integrated and separated contribution to ITG stability are studied both theoretically and numerically.

Here, ZFZF  is presented as an example for clarity of discussion. For the adopted ZFZF-like radial electric field with $\omega\ll\omega_{tr,i}$, we found that the poloidal rotation is the main reason for the significant reduction of the ITG growth rate both in the short- and long-wavelength limit.  In contrast, the up-down symmetric density perturbation, which peaks at the un-favourable curvature region, have weak suppression effect on ITG turbulence in both short- and long-wavelength limit.
The extension of our ITG  stability analysis to include EGAM/GAM with frequency being much higher than ion transit frequency, and thus up-down anti-symmetric density perturbation ($\propto\sin\theta$), is straightforward, since in that case the GAM/EGAM frequency is smaller than ITG growth rate.

The present work, motivated to understand local properties and mechanism of  ITG stability in the existence of a given $E_r$ in ballooning space, found that, the radial electric field always plays a stabilizing role on ITG.   This model can also be applied to study the possible cross-scale interaction between ITG and AEs, with the former being microscopic and the latter being macro- or meso- scale, mediated by ZFZF. In such a two-pray one-predator system, effects of  ZFZF generated by one  turbulence can be considered as ``passive" or ``equilibrium" for another turbulence. Additionally, it is of great interest and importance to investigate the nonlinear modulation of ITG by EGAM/GAM, which can be excited by AEs (internally) and neutral beam injection (externally), and  can act as an active control of ITG turbulence.  But, in contrast to ZFZF, the time scale separation of EGAM/GAM with ITG is not necessarily always satisfied, and depends on specific experimental conditions. Hence, fully nonlinear process, i.e., DW-GAM nonlinear evolution, should be considered to account for comparable frequency between ITG and GAM, and will be investigated in a future publication.

\section*{Acknowledgements}

This work is supported by the National Key R\&D Program of China  under Grant No. 2017YFE0301900, and the National Science Foundation of China under grant No. 11875233.   The original idea of investigating ITG stability in the existence of EGAM induced periodic modification to the potential well was provided by Prof. Liu Chen (Zhejiang University and University of California, Irvine).

\section*{Data Availability}
The data that support the findings of this study are available from the corresponding author upon reasonable request.

\end{document}